\documentclass[twocolumn, prl]{revtex4}
\usepackage{subfigure}
\usepackage{epsfig,amssymb,amsmath,graphicx,enumerate,color}

\usepackage{todonotes}

\DeclareMathAlphabet{\matheu}{U}{eus}{m}{n}

\newcommand{\ket}[1]{|{#1}\rangle}
\newcommand{\braket}[2]{\langle{#1}|{#2}\rangle}
\newcommand{\ketbra}[2]{|{#1}\rangle\!\langle{#2}|}
\newcommand{\e}{{\mathrm e}}

\begin{document}

\title{Achieving minimum-error discrimination of an arbitrary set of laser-light pulses}

\author{Marcus P. da Silva, Saikat Guha, Zachary Dutton}
\affiliation{Quantum Information Processing Group, Raytheon BBN
  Technologies, Cambridge, MA 02138}

\begin{abstract} Laser light is widely used for communication and sensing applications, so
the optimal discrimination of coherent states---the quantum states of light
emitted by an ideal laser---has immense practical importance. However, quantum
mechanics imposes a fundamental limit on how well different coherent states
can be distinguished, even with perfect detectors, and limits such
discrimination to have a finite minimum probability of error.
While conventional optical receivers lead to error rates well above this
fundamental limit, Dolinar found an explicit receiver design involving optical
feedback and photon counting that can achieve the minimum
probability of error for discriminating any two given coherent states. The
generalization of this construction to larger sets of coherent states has
proven to be challenging, evidencing that there may be a limitation inherent
to a linear-optics-based adaptive measurement strategy. In this Letter, we
show how to achieve optimal discrimination of any set of coherent states using
a resource-efficient quantum computer. Our construction leverages a recent
result on discriminating multi-copy quantum hypotheses and
properties of coherent states. Furthermore, our construction is reusable,
composable, and applicable to designing quantum-limited processing of
coherent-state signals to optimize any metric of choice. As illustrative
examples, we analyze the performance of discriminating a ternary alphabet, and
show how the quantum circuit of a receiver designed to discriminate a binary
alphabet can be reused in discriminating multimode hypotheses. Finally, we
show that our result can be used to achieve the quantum limit on the rate
of classical information transmission on a lossy optical
channel, which is known to exceed the Shannon
rate of all conventional optical
receivers.
\end{abstract}

\maketitle
 
\vspace{3mm} 
{\bf Background ---} 
Helstrom provided a set of necessary and sufficient conditions on the
measurement that yields the minimum average probability of error in
discriminating $K \ge 2$ distinct quantum states~\cite{Hel76}. However, for
optical state discrimination, this mathematical specification of measurement
operators does not usually translate into an explicit receiver specification
realizable using standard optical components, thus leaving a gap between the
minimum error probability (the \emph{Helstrom limit}) and the minimum
achievable by conventional measurements, e.g., homodyne, heterodyne, and
direct detection.

Dolinar proposed a receiver that achieves the Helstrom limit exactly
for discriminating any two coherent-state signals
\cite{Kennedy72,Dol73,Cooketal07}. The receiver works by applying one
of two time-varying optical feedback waveforms to the laser pulse
being detected, and instantaneously switching between the two feedback
signals at each click event at a shot-noise-limited photon
counter. More recently~\cite{Tak06}, it has been shown that an
arbitrary multi-mode binary projective measurement can be implemented
using adaptive linear-optic feedback and photon counting, thereby
subsuming the Dolinar receiver.

\begin{figure*}[h!]
\begin{center}
\subfigure[An optimal BPSK receiver~\cite{Tak06}]{
  \includegraphics[width=.65\textwidth]{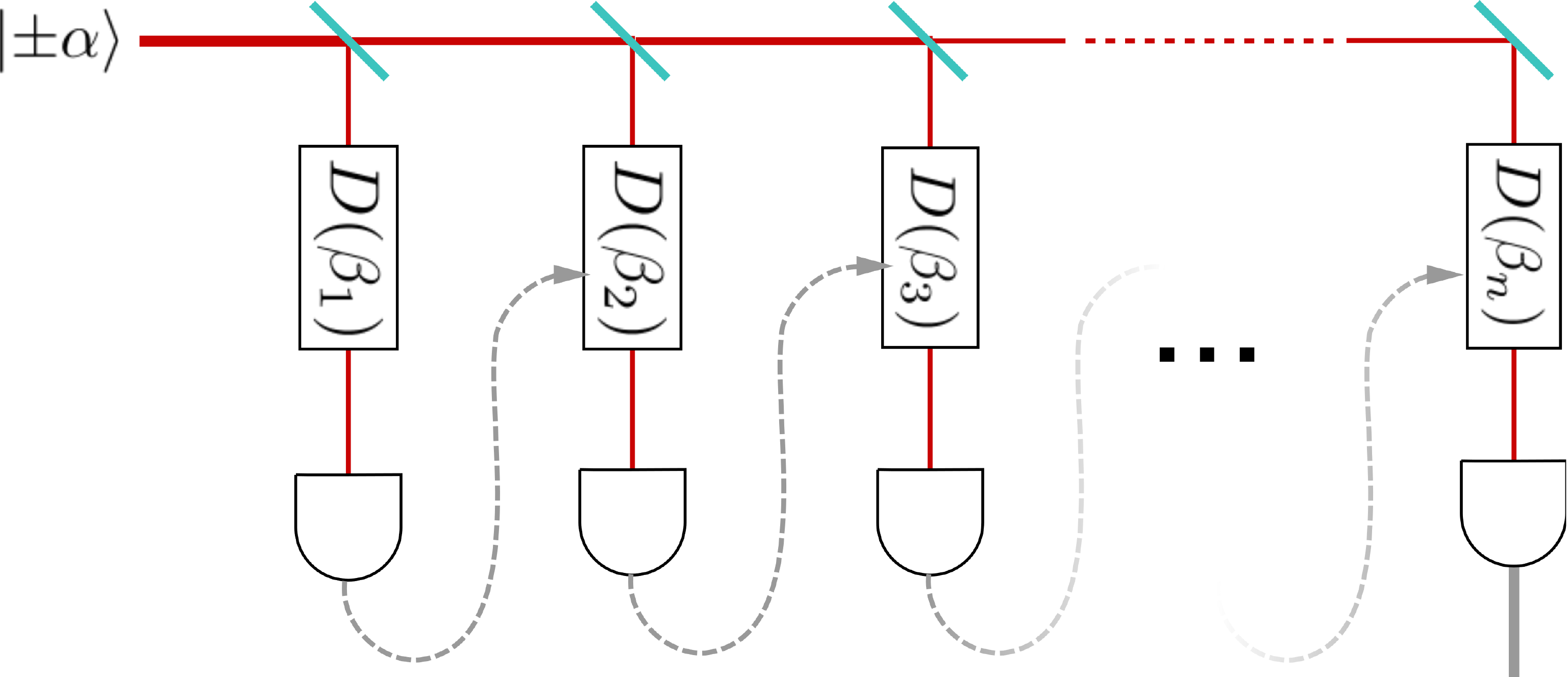}
}
\subfigure[Unitary compression receiver]{
  \includegraphics[width=\textwidth]{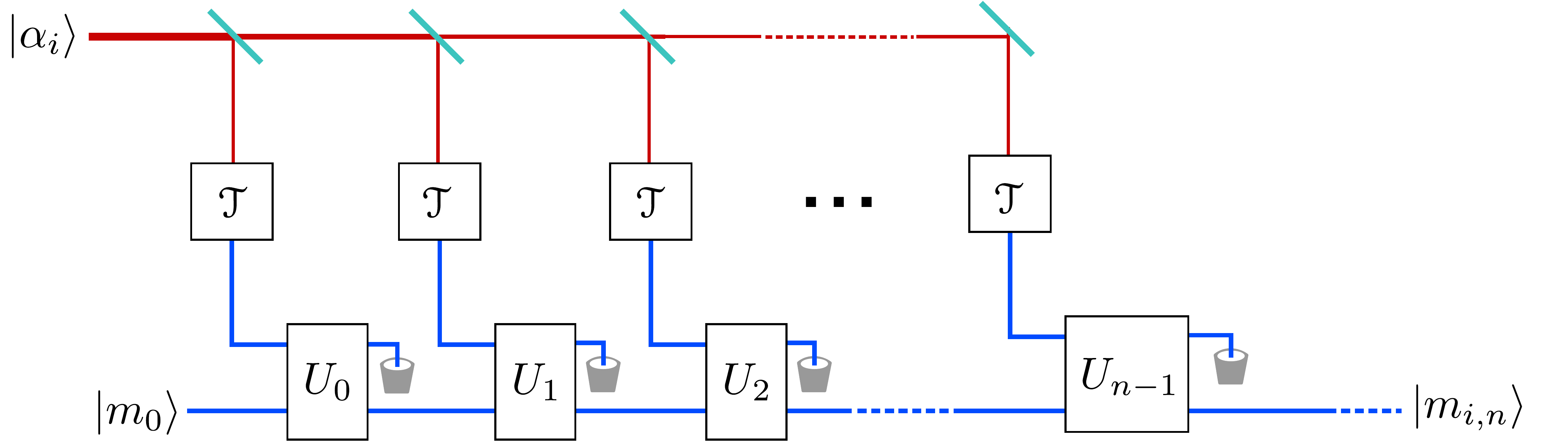}
}
\caption{(a) It is possible to distinguish between two coherent states
  (red lines)
  optimally by ``slicing'' the input state and measuring each slice
  adaptively---the outcome of each photon detection measurement (grey
  dashed arrows) being used to perform a displacement on the input of the
  next measurement. Once the final slice is measured, the final
  outcome is used to make a decision about which hypothesis was
  more likely to have been received. This is the class of receivers 
  shown to be optimal for binary hypotheses in~\cite{Tak06}. 
  (b) Here we demonstrate that,
  instead of measuring each slice adaptively, one can transfer
  ($\matheu{T}$) each
  coherent slice into a qubit (blue lines), and the information in these qubits
  can be efficiently and coherently {\em compressed} by the unitaries
  $U_\ell$ into a small ancilla quantum register~\cite{Blu12}, so that the final state
  $\ket{m_{i,n}}$ of the register can then be measured or processed
  further as part of a multimode receiver as discussed in the
  text. This unitary compression receiver can be customized to
  any set of coherent-state hypotheses, and its design is independent
  of the figure of metric being optimized, as all information about
  the received state is compressed into the final state of the ancilla register.
  \label{fig:circuits}}
\end{center}
\end{figure*}

For discriminating multiple ($K>2$) coherent states, no optical receiver that
achieves the Helstrom bound has been proposed, although several sub-optimal
receivers have been proposed to discriminate more than two coherent-state
signals~\cite{Dolinar83, Bondurant93} and implemented~\cite{Chenetal12},
improving over the performance of conventional receivers. Much like the
receiver in~\cite{Tak06}, each one of these sub-optimal receivers operate via
a common philosophy: that of {\em slicing} a coherent-state pulse into smaller
coherent states, detecting each slice via photon counting after coherent
addition of a local field, and feeding forward the detection outcome to the
processing of the next slice, as illustrated in Fig.~\ref{fig:circuits}(a).

Attaining the quantum-limited channel capacity of an optical channel to carry
classical information---the {\em Holevo limit}---requires joint detection
over long coherent-state codewords~\cite{Guh11}. However, recent work has
shown that even a joint-detection receiver involving the most general
coherent-state feedback, passive linear optics and photon counting {\em
cannot} attain the Holevo limit~\cite{Chu11}. This proves that lasers, passive
linear optics and photon counting (the resources used by the Dolinar receiver)
are not sufficient for general minimum-error optical state discrimination, unlike
the binary-discrimination case.

{\bf Discriminating multiple coherent states ---} Our main result, the
description of an optimal receiver for $K$-ary coherent-state discrimination,
relies on two observations.

The first observation is that quantum-limited performance of any
processing of an ensemble of $K$ linearly independent pure states, ${\cal
  A} = \left\{\ket{\alpha_j}\right\}$, $1 \le j \le K$, is completely
described by the Hermitian Gram matrix $\Gamma$, whose elements are
the inner products $\gamma_{jk} \equiv
\langle{\alpha_j}|\alpha_k\rangle$. Therefore, by a simple
dimensionality argument, there must exist a unitary map $U_{\cal A}$
that maps the state ensemble ${\cal A}$ to an ensemble of states of
$\lceil \log_2K \rceil$ qubits. Note that, even though the coherent
states $|\pm\alpha\rangle = \sum_{n=0}^{\infty}e^{-|\alpha|^2/2}([\pm
\alpha]^n/\sqrt{n!})|n\rangle$ are embedded in an infinite dimensional
space spanned by the photon number (Fock) states
$\left\{|n\rangle\right\}$, since the hypothesis states span a finite
dimensional space, the entire relevant information can be {\em
  compressed} down to a finite dimensional space, e.g., the state of a
collection of qubits.  Helstrom showed that minimum-error
discrimination of $K$ linearly independent pure states in $\cal A$
requires a $K$-outcome projective measurement in the span of $\cal
A$~\cite{Hel76}. Thus, assuming $U_{\cal A}$ can be implemented to
compress the hypothesis states to one of $K$ states of
$\lceil\log_2K\rceil$ qubits, Helstrom's optimal projective
measurement can be implemented by a unitary rotation on the compressed
state, followed by measurement of the qubits.  However, it is not at
all obvious how to implement $U_{\cal A}$ in the coherent state case,
as it corresponds to a highly non-linear optical transformation. This
is where, our second observation---a unique property of coherent
states---comes in handy.

The second observation is that a coherent state $|\alpha_j\rangle$, can be
{\em sliced} into $n$ independent coherent states of smaller amplitudes using
a $1$:$n$ symmetric beam-splitter,
\begin{equation}
\ket{\alpha_j} \to 
\ket{\beta_j}\otimes\ket{\beta_j}\otimes\cdots\otimes\ket{\beta_j},
\end{equation}
$\beta_j=\alpha_j/\sqrt{n}$. Furthermore, for $n$ large, $|\beta_j|
\ll 1$ which implies the average photon number in the slices is small,
and we find
\begin{equation}
\ket{\beta_j} \approx {\ket{0} + \beta_j \ket{1}\over\sqrt{1+|\beta_j|^2}}
\equiv\ket{h_j}.
\end{equation}
In other words, any coherent state can be split into multiple copies
of weak coherent states, each of which can be faithfully represented
by a qubit encoded in the vacuum $\ket{0}$ and single photon $\ket{1}$
states (also known as a {\em single-rail qubit
  encoding}~\cite{Kok07}).

It is now easy to see how to explicitly construct $U_{\cal A}$ by
concatenating $n$ unitary gates. We start with a $\lceil \log_2K
\rceil$-qubit ancilla register prepared in some initial state
$\ket{m_0}$. We then unitarily compress the received state
$\ket{\alpha_j}$, one slice $\ket{\alpha_j/\sqrt{n}}$
at a time, into the state of the register, following the proposal by
Blume-Kohout {\em et al.} for general multi-copy quantum hypothesis
test~\cite{Blu12}. This is done using a sequence of $(1 + \lceil
\log_2K \rceil)$-qubit entangling gates $U_\ell$ acting jointly on the
$\ell^{\rm th}$ slice and the ancilla register, eventually transforming
the ancilla register to a state $|m_{j,n}\rangle$.  More precisely, we
require
\begin{equation}\label{eq:u-constraint}
U_\ell\ket{h_j}\ket{m_{j,\ell}}=\ket{\phi}\ket{m_{j,\ell+1}}, 
\end{equation}
for all hypothesis states $\ket{h_j}$, where $\ket{\phi}$
is a fixed quantum state independent of the hypothesis $i$ which can
then be discarded as it contains no relevant information.

It is possible to do this state transfer in such a way that the Gram
matrix elements of the compressed ensemble approaches the Gram matrix
of the original coherent-state ensemble, i.e., $\gamma^{(n)}_{jk}
\equiv \langle{m_{j,n}}|{m_{k,n}}\rangle \to \gamma_{jk}$, in the
limit of many slices ($n \to \infty$).  This follows from the fact
that the collection of qubit states approximating the slices of
coherent states have inner-products arbitrarily close to the
overlap of the original coherent states in the limit of large $n$, i.e.  $
\lim_{n\to\infty}
\langle h_j|h_k \rangle^n
=\braket{\alpha_j}{\alpha_k} $.  As we discussed above, once the
coherent-state ensemble $\cal A$ has been compressed faithfully into
the ancilla register, any additional quantum processing can be
performed without any loss in performance.  For the purposes of
minimum error discrimination between the hypothesis states, a
projective measurement of the ancilla states is required. This can be
implemented by a unitary rotation $U_H$ on the register's state
followed by a computational-basis measurement. Since the number of
gates necessary to implement an arbitrary $N$ qubit unitary is
exponential in $N$~\cite{Kit97}, the number of gates necessary to
build $U_H$ is only polynomial in the number of hypotheses
$K$. Therefore, the receiver is efficient from the resource-scaling
point of view. Moreover, computing $U_H$ requires solving a set of
$K^2$ non-linear simultaneous equations prescribed by
Helstrom~\cite{Hel76}, which can be done with an overhead that is
polynomial in $K$. See Fig.~\ref{fig:circuits}(b) for an illustration
of the compression unitaries acting sequentially on $n$ slices of the
received state.

Finally, unlike Dolinar's optimal binary-discrimination
receiver---which adaptively, albeit {\em destructively}, measures tiny
slices of the received coherent state---our receiver coherently
couples slices of the received coherent state into a
$\lceil\log_2K\rceil$ dimensional quantum register, the final state of
which has the entirety of the relevant quantum information that was
present in the original coherent state. The adaptive destructive
measurement strategy corresponds to a local operations and classical
communication (LOCC) strategy. If our final measurement is writen in
terms of its action on the slices, it becomes clear that our approach
amounts to a collective quantum measurement, thereby sidestepping the
LOCC limitations of the generalized Dolinar strategy.

\vspace{3mm}

{\bf Binary discrimination ---} As an illustration, consider the
compression unitaries for distinguishing between the equi-prior BPSK
ensemble of coherent states, $\ket{\pm\alpha}$ where $\alpha \in
{\mathbb R}$. Any pair of coherent states can be transformed to this
ensemble via simple linear-optical transformations. Although the Dolinar receiver
can distinguish these states with the quantum-limited minimum
probability of error, the compression approach is more flexible, as it
is possible to perform additional quantum processing of the compressed
state in the ancilla register, enabling multimode applications, as will be
discussed later. Moreover, the construction of the
minimum-error-discrimination receiver for the binary case generalizes
straightforwardly to larger sets of hypotheses.

After slicing via a symmetric $1$:$n$ beamsplitter, the hypothesis states can be
approximated well by the states
\begin{equation}
\ket{h_j}^{\otimes n} = \left({\ket{0} + \beta_j
\ket{1}\over\sqrt{1+|\beta_j|^2}}\right)^{\otimes n}
\end{equation}
where $\beta_{\pm1}=\pm\alpha/\sqrt{n}$. We assume that $n$ is chosen large
enough such that $|\beta_j|\ll 1$ holds. Since $K=2$, only $1$ ancilla qubit
is required. Let us denote the input state of the ancilla qubit for the
$\ell^{\rm th}$ compression step under hypothesis $j$ as $\ket{m_{j,\ell}}$ (see
Fig.~\ref{fig:circuits}(b)), the initial state of the ancilla
$\ket{m_{j,0}}=\ket{0}$, and the $\ell^{\rm th}$ compression unitary, $U_\ell$. The $0^{\rm
th}$ compression step is to map $\ket{h_j}\ket{m_{j,0}}$ to
$\ket{0}\ket{m_{j,1}}$, and a natural choice to make is
$\ket{m_{j,1}}=\ket{h_j}$, so that $U_0$ is just the exchange of the input and
the ancilla states---the well-known two-qubit swap gate.

The subsequent compression operations have to satisfy
\begin{align}\label{eq:bpsk-unitary}
U_\ell \ket{h_{-1}}\ket{m_{-1,\ell}} &= \ket{0}\ket{m_{-1,\ell+1}},\\
U_\ell \ket{h_{+1}}\ket{m_{+1,\ell}} &= \ket{0}\ket{m_{+1,\ell+1}},
\end{align}
i.e., all the information about the received slices is compressed,
sequentially, into the ancilla qubit. We choose $\ket{m_{j,\ell}}$ to be of the
same general form as $\ket{h_{j}}$ but with different parameters. In
particular, we can choose
\begin{equation}
\ket{m_{j,\ell}} = {\ket{0} + j B_{(\ell)}
\ket{1}\over\sqrt{1+B_{(\ell)}^2}},
\end{equation}
for some $B_{(\ell)}$, and given the fact that $U_\ell$ preserves the inner product,
we obtain the recursion relation $B_{(\ell+1)}=\sqrt{\alpha^2/n+B_{(\ell)}^2\over
1+\alpha^2 B_{(\ell)}^2/n}$, and $B_{(0)}=0$. This can be solved analytically and
the parameterization for the $U_\ell$ immediately follows (see Appendix). It can
be easily shown that the measurement that distinguishes the final compressed
hypotheses $\ket{m_{i,n-1}}$ with minimum error probability is a measurement
in the basis $\ket{0}\pm\ket{1}$. The error probability is given by
$(1-B_{(n)})^2/2(1+B_{(n)}^2)$, which in the limit of $n\to\infty$, equals the
Helstrom bound $(1-\sqrt{1-{\mathrm e}^{-4\alpha^2}})/2$.

\vspace{3mm} {\bf Transferring optical states to qubits ---} As
described above, the ability to implement a universal set of qubit
operations is essential for the implementation of our proposal for
quantum-limited minimum-error discrimination of coherent-state
signals. However, operations on single-rail qubits suffer from
significant technical challenges~\cite{Kok07}. Viable optical
implementations of general deterministic single-qubit operations
remain unknown, and two-qubit interactions are also challenging.

\begin{figure*}[h!]
\begin{center}
\subfigure[BPSK alphabet, ideal unitary transfer]{
  \includegraphics[width=.45\textwidth]{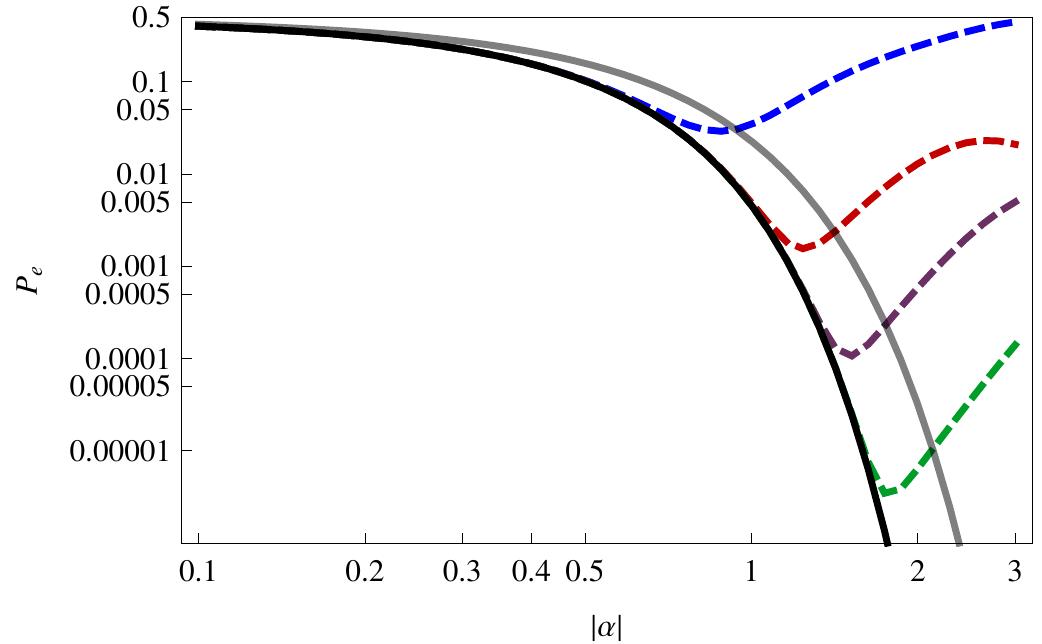}
}
\subfigure[BPSK alphabet, STIRAP transfer]{
  \includegraphics[width=.45\textwidth]{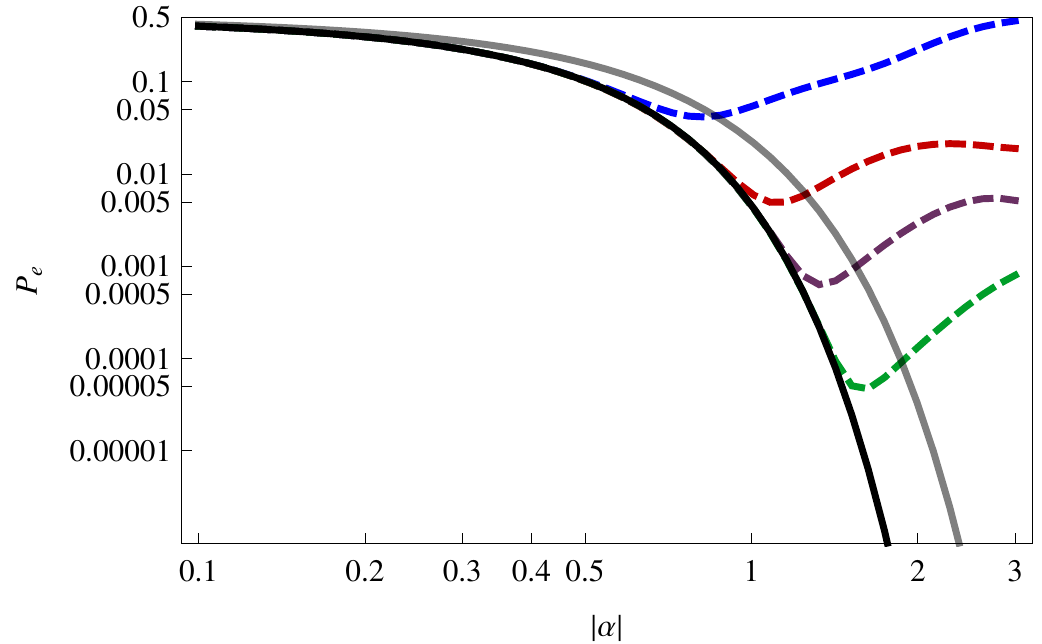}
}
\subfigure[3ASK alphabet, ideal unitary transfer]{
  \includegraphics[width=.45\textwidth]{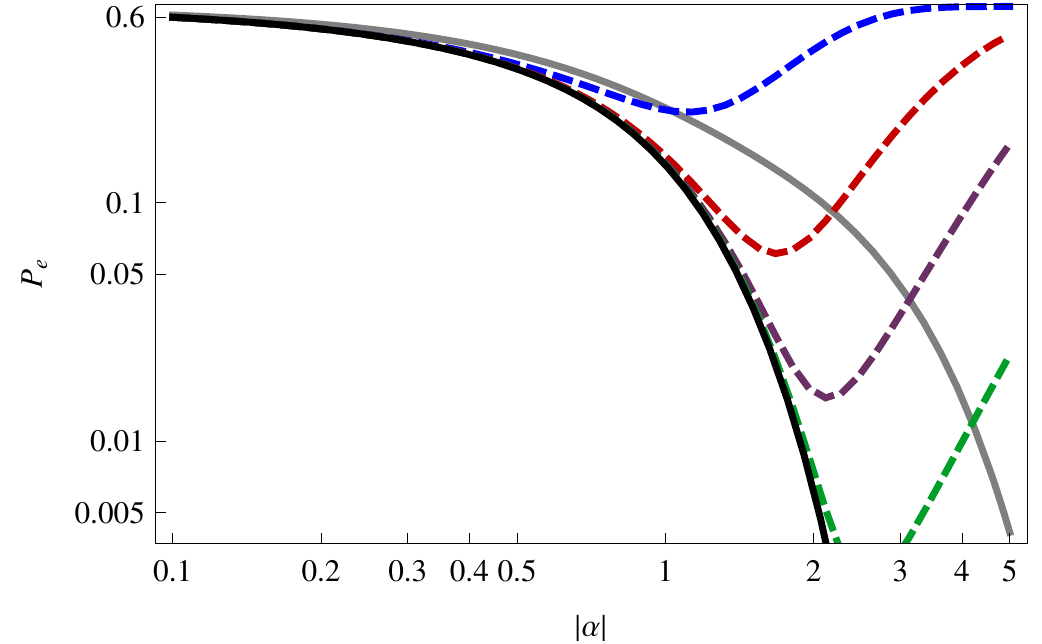}
}
\subfigure[3ASK alphabet, STIRAP transfer]{
  \includegraphics[width=.45\textwidth]{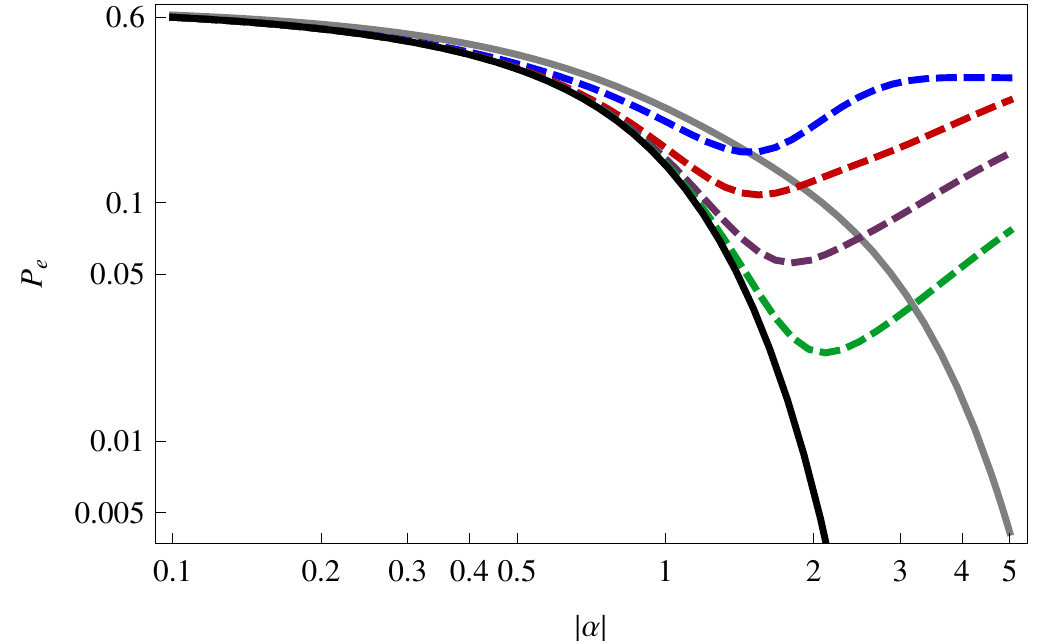}
}
\caption{
The minimum error probability of distinguishing between a set of coherent
states as a function of the amplitude $\alpha$ follows the well known bound
given by Helstrom~\cite{Hel76} (all black curves). Maximum-likelihood
estimation of the hypothesis based on homodyne detection leads to higher error
rates (all grey curves). The receiver described here has a probability of
error arbitrarily close to the minimum for sufficiently small
$\alpha/\sqrt{n}$. This is illustrated by the dashed curves for ideal unitary
transfer (left column) and STIRAP state transfer (right column) both in the
binary alphabet case (top row) as well as in the ternary alphabet case (bottom
row), for $n=2$ (blue), $n=10$ (red), $n=30$ (purple), and $n=100$ (green)
slices.
\label{fig:bpsk-3ask-prob-error}}
\end{center}
\end{figure*}

This problem can be avoided by transfering the optical state of the
slices into another qubit realization where such operations are more
easily implemented. We consider two possibilities: (a) the ideal
mapping of the $\{\ket{0},\ket{1}\}$ subspace of the
optical mode to a qubit, and (b) a stimulated Raman adiabatic passage
(STIRAP) based transfer of the optical mode state to an atom-like
system that can then be manipulated as a qubit~\cite{Oi12}. In both
cases we take the unitary manipulation of the resulting states to be
unrestricted, and consider only the degradation effects of the
transfer process. The result is that, instead of having each slice of
the coherent state map to the pure states $\ket{h_j}$, one obtains the
mixed states,
\begin{align}
\rho_{\beta_j}^{\textrm{(a)}} &=\e^{-\beta_j^2}\ketbra{h_j}{h_j} +
(1-\e^{-\beta_j^2})\ketbra{0}{0},\\ 
\rho_{\beta_j}^{\textrm{(b)}} &= \e^{-\beta_j^2}\ketbra{h_j}{h_j} +
(1-\e^{-\beta_j^2})\ketbra{1}{1}, 
\end{align}
corresponding to the two possibilities for the transfer procedure described
above, and which deviate from the qubit hypotheses due to populations in the
subspace with $2$ photons or more. Despite this source of decoherence, it is
always possible to make the distinguishability between the collection of
transferred states and the collection of corresponding pure state hypotheses
arbitrarily small by making $n$ sufficiently large (see Appendix). This
translates to being able to use the unitaries $U_\ell$ designed for the pure
states $\ket{h_j}$ on the mixed states $\rho_{\beta_j}^{\textrm{(a)}}$ and
$\rho_{\beta_j}^{\textrm{(b)}}$, while remaining arbitrarily close to the
Helstrom bound for the minimum probability of error, as illustrated in
Figure~\ref{fig:bpsk-3ask-prob-error}.

\vspace{3mm}
{\bf Ternary and multimode cases ---} 
Let us consider now the case of 3 coherent-state hypotheses, which for
simplicity we take to be $\ket{j\alpha}$ for $j\in\{-1,0,1\}$ (i.e. a
displaced ternary amplitude-shift keying or 3ASK alphabet, where the
displacement adds symmetry to simplify the analysis). No known generalization
of the optimal receivers for BPSK can achieve the minimum probability of error
in distinguishing these 3 states, but our coherent compression receiver can.
Using the same approach of slicing the coherent states to obtain states that
are well approximated by qubit states, we simply need to specify the unitaries
$V_\ell$ that perform the compression. In this case, we choose the ancilla states
to have the form
\begin{equation}
\ket{m'_{j,\ell}} = 
{\ket{00}+j C_{(\ell)}\ket{01}+ j^2 D_{(\ell)}\ket{11}\over\sqrt{1+j^2C_{(\ell)}^2+j^2D_{(\ell)}^2}},
\end{equation}
with the resulting coupled recurrence relations given by
\begin{align}
C_{(\ell+1)} &= {\alpha\over\sqrt n} \sqrt{C_{(\ell)}^2+\alpha^2/n+\alpha^2 D_{(\ell)}^2/n}\, ,\\
D_{(\ell+1)} &= \sqrt{D_{(\ell)}^2+\alpha^2 C_{(\ell)}^2/n}\,\\
C_{(0)} &=D_{(0)}=0,
\end{align}
which, as in the binary case, can be solved exactly. The relevant
parameters for $V_\ell$ follow straightforwardly from a set of
constraints analogous to equations~\eqref{eq:u-constraint}
and~\eqref{eq:bpsk-unitary} (see Appendix).

For a finite number of slices $n$, the minimal error
probability for distinguishing the states in the 3ASK alphabet can be
computed semi-analytically (see Appendix), and the same general
behavior of the BPSK case, where the performance approaches the
Helstrom limit, can be observed for the 3ASK case as well, as
illustrated in Fig.~\ref{fig:bpsk-3ask-prob-error}(c) and (d).

This compression receiver approach opens the door to quantum-limited
optimization of metrics other than the probability of error, in
detection of a coherent-state ensemble. Examples are maximizing the
one-shot accessible information of an ensemble for the optimal
measurement choice~\cite{Hol73}, and minimizing the phase-space
Euclidean norm in detecting a coherent state from a constellation.
This is because the compression operations are independent of how the
final measurement of the ancilla register is optimized.

Another notable feature of this general class of receivers, as alluded to
earlier, is the possibility of additional quantum processing of the compressed
states. This enables, in particular, the discrimination of multimode coherent
states (such as the ones used in sensing or coded communications applications)
by reusing the compression operations for single mode states. For example, if
one would like to discriminate between the $3$ states
\begin{gather}
\ket{+\alpha}\ket{+\alpha}\ket{-\alpha}\ket{-\alpha},\notag\\
\ket{-\alpha}\ket{+\alpha}\ket{+\alpha}\ket{-\alpha},\notag\\
\ket{-\alpha}\ket{-\alpha}\ket{+\alpha}\ket{+\alpha},\notag
\end{gather}
one would slice and compress the state of each mode independently using our
coherent BPSK receiver, but instead of measuring the four compressed
ancilla states corresponding to each mode, one would perform
additional compression of these states into a three dimensional
subspace of a 2 qubit register. The coherent compression ensures that
the Helstrom bound for distinguishing between the three states can be
reached, while the individual measurement of each mode leads to a
higher error probability.

This modularity also allows for a systematic way to develop receivers
that surpass the Shannon capacity of an optical channel with
structured optical receivers, and approach the Holevo limit to the
rate at which classical data can be sent reliably over an optical
channel~\cite{Hol73, Guh11}. This can be done by first compressing
each of the $N$ codeword symbols---which are coherent states from a
known and finite $M$-ary constellation---into the states of $N$
separate $\lceil\log_2M \rceil$-qubit registers, which are then unitarily
compressed into a $O(rN)$ qubit state, where $r$ is the rate of
the code, in order to perform minimum-error discrimination of the
codeword states. One may be able to find efficient implementations of
this unitary compression by leveraging the structure of
Holevo-capacity-achieving error-correcting codes~\cite{GuhWil12}.

\vspace{3mm} {\bf Summary ---} We have demonstrated how to construct a
receiver to discriminate between any set of coherent-state (laser
light) signals, using a small special-purpose quantum computer. Our
solution leverages two properties of coherent states of light---first,
that splitting a coherent state via a beamsplitter produces
independent coherent states with smaller amplitudes, and second, that
a coherent state of a small amplitude is well approximated by a qubit
encoded by the vacuum and the one-photon Fock state of an optical
mode, the single-rail qubit. These two properties, in conjunction with
a recent result on distinguishing multi-copy quantum hypotheses by a
sequential coherent-processing receiver~\cite{Blu12}, leads to an
explicit construction of an optimal receiver for any specified
coherent-state hypothesis test. This solves the long standing problem
of building receivers that can achieve the Helstrom limit for
non-binary coherent-state signals. There remain practical challenges
in implementing our receiver because universal quantum computing on
single-rail qubits still eludes us. However, we address one
potentially practical means to implement our receiver by transferring
the state of an optical mode to an atomic system~\cite{Oi12}.  Our
solution also opens the door to the quantum-limited optimization of
other important metrics, such as achieving the Holevo bound on the
rate of classical information transmission over a quantum
channel. This would allow for reliable information transmission on an
optical channel at rates higher than the Shannon limit of any known
receiver.

\section{Acknowledgment}

The authors thank Mark Wilde for pointing out~\cite{Blu12} and for
useful discussions. This research was supported by the DARPA
Information in a Photon (InPho) program under contract number
HR0011-10-C-0162.

\appendix

\section{Appendix}

\section{Ideal unitary transfer of $0/1$ photon subspace to qubits}

Given any quantum state of an optical mode, decomposed in the basis of
Fock states (or photon number states) $\ket{n}_F$ for $n\ge0$, it is
in principle possible to unitariy swap the contents of the
$\ket{0}_F/\ket{1}_F$ subspace with a qubit. The unitary that performs
this operation is
\begin{multline}
\ketbra{0}{0}_F\otimes\ketbra{0}{0}
+\ketbra{1}{1}_F\otimes\ketbra{1}{1}
+\ketbra{0}{1}_F\otimes\ketbra{1}{0}\\
+\ketbra{1}{0}_F\otimes\ketbra{0}{1},
\end{multline}
and by applying this operation to the state $\ket{\beta_j}\ket{0}$ and
tracing out the optical mode, one obtains
\begin{equation}
\rho_{\beta_j}^{\textrm{(a)}} =\e^{-\beta_j^2}\ketbra{h_j}{h_j} +
(1-\e^{-\beta_j^2})\ketbra{0}{0},
\end{equation}
so the fidelity to $\ket{h_j}$ is given by
\begin{equation}
F^{\textrm{(a)}}(\beta_j) = \e^{-|\beta_j|^2}|\beta_j|^2 + {1\over1+|\beta_j|^2}.
\end{equation}
One finds
\begin{equation}
F^{\textrm{(a)}}(\alpha/\sqrt{n})^n = 1 - {|\alpha|^6\over2n^2} +
O\left({|\alpha|^8\over n^3}\right).
\end{equation}
so that even by manipulating only the $0/1$ photon subspace and
ignoring the higher excitations, the collection of transferred states
can be made arbitrarily close to the collection of ideal slices of any
hypothesis.

\section{STIRAP transfer of $0/1$ photon subspace to qubits}

The transfer of the photonic excitation into a cavity mode enables the
use of stimulated Raman adiabatic passage (STIRAP) to coherently
exchange a single excitation between the cavity mode and a qubit. The
unitary that corresponds to this operation is
\begin{multline}
\sum_{n=1}^\infty\ketbra{n-1}{n}_F\otimes\ketbra{1}{0}
+\sum_{n=1}^\infty\ketbra{n}{n-1}_F\otimes\ketbra{0}{1}\\
+\ketbra{0}{0}_F\otimes\ketbra{0}{0}
\end{multline}
and by applying this operation to the state $\ket{\beta_j}\ket{0}$ and
tracing out the optical mode, one obtains
\begin{equation}
\rho_{\beta_j}^{\textrm{(b)}} =\e^{-\beta_j^2}\ketbra{h_j}{h_j} +
(1-\e^{-\beta_j^2})\ketbra{1}{1}
\end{equation}
so the fidelity to $\ket{h_j}$ is given by
\begin{equation}
F^{\textrm{(b)}}(\beta_j) = \e^{-|\beta_j|^2} + {|\beta_j|^2\over1+|\beta_j|^2}.
\end{equation}
One finds a similar series expansion for the sliced case
\begin{equation}
F^{\textrm{(b)}}(\alpha/\sqrt{n})^n 
= 1 - {|\alpha|^4\over2n} +
O\left({|\alpha|^6\over n^2}\right).
\end{equation}
so that just as in the ideal unitary transfer case, the collection of
transferred states can be made arbitrarily close to the collection of
ideal slices of any hypothesis.

Similar performance can also be obtained via a tunable Jaynes-Cummings
interaction between a cavity mode and qubit-like system.

\section{Compression unitary for BPSK alphabet}

The recursion relation for $B_{(\ell)}$ can be solved by using standard
methods, yielding
\begin{equation}
B_{(\ell)}=\sqrt{1+{2\over{(-1)^\ell\left({\beta^2+1\over\beta^2-1}\right)^{\ell+1}-1}}}
\end{equation}
where $\ell>0$ and $\beta$ is the amplitude of the hypothesis slices.

Given this parameter, and the constraints
\begin{align}
U_\ell \ket{h_{-1}}\ket{m_{-1,\ell}} &= \ket{0}\ket{m_{-1,\ell+1}},\notag\\
U_\ell \ket{h_{+1}}\ket{m_{+1,\ell}} &= \ket{0}\ket{m_{+1,\ell+1}},\notag
\end{align}
the relevant 2 dimensional block of $U_\ell$ is uniquely determined.  For
our purposes, the remaining block can be chosen arbitrarily to
complete the unitary, resulting in
\begin{align}
U_{\ell>0} = \left[
\begin{array}{cccc}
{1\over\sqrt{1+\beta^2 B_{(\ell)}^2}} & 0 & 0 & {\beta B_{(\ell)}\over\sqrt{1+\beta^2
  B_{(\ell)}^2}}\\
0 & * & * & 0\\
0 & * & * & 0\\
{\beta B_{(\ell)}\over\sqrt{1+\beta^2 B_{(\ell)}^2}} & 0 & 0 & -{1\over\sqrt{1+\beta^2 B_{(\ell)}^2}}
\end{array}
\right],
\end{align}
where the $*$ denote free parameters.

Finally, the inner product between compressed ancilla states for the
hypothesis states
$\ket{h_j}=
{\ket{0}+j{\alpha\over\sqrt{n}}\ket{1}\over\sqrt{1+{\alpha^2\over n}}}$ 
is given by
\begin{equation}
\left(1-{\alpha^2\over n}\over 1+{\alpha^2\over n}\right)^{n+1}
\end{equation}
which in the limit of $n\to\infty$ reduces to
$\braket{-\alpha}{\alpha}=\e^{-2\alpha^2}$ as claimed.

\section{Compression unitary for 3ASK alphabet}

The recursion relation for $C_{(\ell)}$ and $D_{(\ell)}$ can also be solved
analytically, yielding
\begin{align}
C_{(\ell)}&=\sqrt{(1+\beta^2)^\ell-(1-\beta^2)^\ell\over2}\\
D_{(\ell)}&=\sqrt{{(1+\beta^2)^\ell+(1-\beta^2)^\ell\over2}-1}
\end{align}
where $\ell>0$ and $\beta$ is the amplitude of the hypothesis slices.

The constraints for the compression unitary $V_\ell$ have a similar form as in
the BPSK case, although now there are 3 hypotheses, resulting in
\begin{align}\label{eq:3ask-unitary}
V_\ell \ket{h'_{-1}}\ket{m'_{-1,\ell}} &= \ket{0}\ket{m'_{-1,\ell+1}},\\
V_\ell \ket{h'_{0}}\ket{m'_{0,\ell}} &= \ket{0}\ket{m'_{0,\ell+1}},\\
V_\ell \ket{h'_{+1}}\ket{m'_{+1,\ell}} &= \ket{0}\ket{m'_{+1,\ell+1}},
\end{align}
while the remaining degrees of freedom of $V_\ell$ can be
completed arbitrarily for our purposes.
\begin{widetext}
This results in
\begin{align}
V_{\ell>0} = 
\left[
\begin{array}{cccccccc}
1 & 0 & 0 & 0 & 0 & 0 & 0 & 0\\
0 & {C_{(\ell)}\over\sqrt{\beta^2+C_{(\ell)}^2+\beta^2D_{(\ell)}^2}} & 0 & 0 & 
{\beta\over\sqrt{\beta^2+C_{(\ell)}^2+\beta^2D_{(\ell)}^2}} & 0 & 0 & 
{\beta D_{(\ell)}\over\sqrt{\beta^2+C_{(\ell)}^2+\beta^2D_{(\ell)}^2}}\\
0 & * & * & * & * & * & * & *\\
0 & 0 & 
{D_{(\ell)}\over\sqrt{D_{(\ell)}^2+\beta^2C_{(\ell)}^2}} & 0 & 
{\beta C_{(\ell)}\over\sqrt{D_{(\ell)}^2+\beta^2C_{(\ell)}^2}} & 0 & 0 & 0\\
0 & * & * & * & * & * & * & *\\
0 & * & * & * & * & * & * & *\\
0 & * & * & * & * & * & * & *\\
0 & * & * & * & * & * & * & *
\end{array}
\right],
\end{align}
\end{widetext}
where once again the $*$ entries are free parameters.

The Gram matrix (the matrix of inner products) for the 3 compressed
hypothesis states is
\begin{align}
\left[
\begin{array}{ccc}
1 & \left(n\over n+\alpha^2\right)^{n/2} & \left(n\over n+\alpha^2\right)^{n/2}\\
\left(n\over n+\alpha^2\right)^{n/2} & 1 & \left(n-\alpha^2\over n+\alpha^2\right)^n\\
\left(n\over n+\alpha^2\right)^{n/2} & \left(n-\alpha^2\over n+\alpha^2\right)^n & 1
\end{array}
\right],
\end{align}
which reduces to the Gram matrix for the 3 coherent-state hypotheses
in the 3ASK alphabet in the limit of $n\to\infty$ as claimed.

\section{Minimal probability of error measurement for 3 hypotheses
  with isoceles configuration}

Here we briefly describe a minor generalization of an example from
Helstrom's book~\cite{Hel76} so that it can be applied both to the
minimal error probability discrimination of coherent states as well as
qubit hypothesis states (details of the derivation can be found
in~\cite{Hel76}).  

Consider 3 linearly independent pure state
hypotheses with Gram matrix
\begin{align}
\Gamma = \left[
\begin{array}{ccc}
1 & x & x\\
x & 1 & y\\
x & y & 1
\end{array}
\right],
\end{align}
where, in our case, $x=\braket{0}{\alpha}$ and
$y=\braket{-\alpha}{\alpha}$, but this need not be the case in
general. Due to the symmetry of the Gram matrix, the matrix $\Xi$ of
inner products $\braket{w_i}{\psi_j}$ between the optimal projectors
$\ket{w_i}$ and the hypotheses $\ket{\psi_j}$ is restricted to take
the form
\begin{align}
\Xi = \left[
\begin{array}{ccc}
a & d & d\\
d & c & e\\
d & e & c
\end{array}
\right],
\end{align}
where the matrix elements must satisfy the constraints~\cite{Hel76}
\begin{align}
a^2+2b^2&=1,\\
d^2+c^2+e^2&=1,\\
ad+b(c+e)&=x,\\
d^2+2ce&=y,\\
ab&=cd.
\end{align}
Solving the first 4 constraints in terms of $d$ results in
\begin{align}
a&={2 d x + \sqrt{(y-2x^2+1) (y-2d^2+1)}\over1+y},\\
b&={x-2d^2x+xy -d \sqrt{(y-2x^2+1) (y-2d^2+1)}\over(1+y)\sqrt{y-2d^2+1}},\\
c&={1\over2}(\sqrt{y-2d^2+1}+\sqrt{1-y}),\\
e&={1\over2}(\sqrt{y-2d^2+1}-\sqrt{1-y}),
\end{align}
which can then be plugged into the 5th constraint to solve for $d$
numerically. The resulting $\Xi$ matrix describes the optimal
projectors to discriminate between the hypotheses described by the
Gram matrix $\Gamma$ with a minimal probability of error $1-{1\over3}(a^2+2c^2)$.


\begin{thebibliography}{99}

\bibitem{Hel76} C. W.~Helstrom, \emph{Quantum Detection and Estimation
    Theory} (Academic Press, New York 1976).

\bibitem{Dol73} S. J. ~Dolinar,  Res. Lab. of Elec. MIT Tech. Rep. No. 111 (1973).

\bibitem{Blu12} R. Blume-Kohout, S. Croke, M. Zwolak, 
  arXiv:1201.6625 (2012)

\bibitem{Hol73} A. S. Holevo, 
  Teoret. Mat. Fiz. {\bf 17}(3), 319--326 (1973).

\bibitem{GGL+04} V. Giovannetti, S. Guha, S. Lloyd, L. Maccone,
  J. H. Shapiro, and H. P. Yuen, Phys. Rev. Lett. {\bf 92}, 027902
  (2004).

\bibitem{Sha48} C.~E.~Shannon, Bell System Technical Journal, 27,
  pp. 379–423 \& 623–656 (1948).

\bibitem{Guh11} S. Guha, Phys. Rev. Lett. 106, 240502 (2011).

\bibitem{GuhWil12} S. Guha, M. Wilde, arXiv:1202.0533v2 [cs.IT],
  Proc. of IEEE Int. Symp. Inf. Th., MIT, Cambridge (2012).

\bibitem{Kennedy72} R. S.~Kennedy, Res. Lab. of Elec. MIT Tech. Rep. No. 110 (1972).

\bibitem{Cooketal07} R. L. ~Cook, P. J. ~Martin, J. M. ~Geremia,
  Nature, {\bf 446}, April 12 (2007).

\bibitem{Tak06} M. Takeoka, M. Sasaki, and Norbert L{\"u}tkenhaus,
  Phys. Rev. Lett. {\bf 97}, 040502 (2006).

\bibitem{Dolinar83} S. J. Dolinar, Jr.,  Telecom. and Data Acquisition Prog. Re
p. 42 72: 1982; NASA: Pasadena, CA,
(1983).

\bibitem{Bondurant93} R. S.~Bondurant, Opt. Lett. \textbf{18}, 1896 (1993).

\bibitem{Chenetal12} J.~Chen, J. L.~Habif, Z.~Dutton, R.~Lazarus and
  S.~Guha, Nature Photonics \textbf{6} 374 (2012).

\bibitem{Chu11} H.-W. Chung, S. Guha, L. Zheng, Proc. of IEEE
  Int. Symp. Inf. Th. (2011).

\bibitem{Kit97} A. Yu. Kitaev, Russ. Math. Surv. {\bf 52}(6),
  1191--1249 (1997)

\bibitem{Kok07} P. Kok, W. J. Munro, K. Nemoto, T. C. Ralph,
  J. P. Dowling, G. J. Milburn, Rev. Mod. Phys. {\bf 79}, 135--174
  (2007).

\bibitem{Oi12} D. K. L. Oi, V. Potocek, J. Jeffers, arXiv:1207.3011
  (2012).
\end{thebibliography}
\end{document}